\documentclass[runningheads]{llncs}

\usepackage{bbding}
\usepackage{graphicx}
\usepackage{amsmath,amssymb}
\usepackage{booktabs}
\usepackage{multirow}
\usepackage{placeins}
\usepackage{float}
\usepackage{xcolor}
\usepackage{tikz}
\usetikzlibrary{arrows.meta, positioning, calc, fit}
\usepackage[hidelinks]{hyperref}

\makeatletter
\renewcommand\paragraph{\@startsection{paragraph}{4}{\z@}%
  {-12\p@ \@plus -4\p@ \@minus -4\p@}{-0.5em}{\normalfont\normalsize\bfseries}}
\makeatother
\newcommand{\A}{\mathcal{A}}
\newcommand{\Lsup}{\mathcal{L}_{\mathrm{sup}}}
\newcommand{\Lssl}{\mathcal{L}_{\mathrm{ssl}}}
\newcommand{\Lksp}{\mathcal{L}_{\mathrm{ksp}}}
\newcommand{\Lang}{\mathcal{L}_{\mathrm{ang}}}
\newcommand{\Lvel}{\mathcal{L}_{\mathrm{vel}}}
\newcommand{\Ljssl}{\mathcal{L}^{\mathrm{JSSL}}}
\newcommand{\xgt}{x^{\mathrm{gt}}}
\newcommand{\xhat}{\hat{x}}

\begin{document}

\title{Joint Supervised and Self-Supervised Training with Acquisition-Robust
Techniques for Accelerated 4D Flow MRI Reconstruction}
\titlerunning{JSSL for Accelerated 4D Flow MRI}

\author{Mengyuan Xue\inst{1}\textsuperscript{*}\Envelope \and Bochun Mei\inst{2}\textsuperscript{*}\Envelope}
\authorrunning{M. Xue and B. Mei}
\institute{Department of Pediatrics, University of Texas Southwestern Medical Center,
Dallas, TX, USA\\ \email{mengyuan.xue@utsouthwestern.edu}
\and
Department of Radiology and Biomedical Imaging, University of California, San Francisco,
San Francisco, CA, USA\\ \email{bochun.mei@ucsf.edu}}

\maketitle
\renewcommand{\thefootnote}{*}
\footnotetext{The authors contributed equally to this work.}
\renewcommand{\thefootnote}{\arabic{footnote}}
\setcounter{footnote}{0}

\begin{abstract}
4D flow MRI measures time-resolved, three-directional blood velocity but requires
long acquisition times, and its diagnostic signal is carried by the phase difference
\emph{between} velocity encodings, not by image magnitude. Recent work has developed a per-encoding variational network to 
address image reconstruction in this field. In this work, we incorporate a joint supervised and self-supervised training 
regime and utilize both magnitude and velocity data during supervision. At the same time, we add multiple 
acquisition-robust and conditioning strategies based on the acceleration factors. On the CMRx4DFlow~2026 aortic dataset 
(1.5 and 3\,T), our model lowers RelErr by $38$--$50\%$ and AngErr by $7.0$--$8.7^\circ$ against a training-matched baseline 
across $R{=}10$--$50$, improving on every held-out subject at every acceleration. Our model also shows strong generalization 
ability to transfer on out-of-distribution data by employing the joint training scheme, with an increase of $7.2\%$ in SSIM 
and decrease of $38\%$ and $31\%$ in AngErr and RelErr respectively.

\end{abstract}

\section{Introduction}
4D flow MRI non-invasively encodes time-resolved three-dimensional velocity across a volume, enabling quantification of 
hemodynamic biomarkers throughout the heart and great vessels~\cite{markl2012fourdflow}. Its clinical adoption is limited 
by long acquisition and reconstruction times, requiring prospective undersampling. The CMRx4DFlow~2026 challenge targets this 
regime with over 400 clinical cases from multiple centers, three field strengths (1.5/3/5\,T), multiple vendors, and 3D 
Cartesian k-t~Gaussian undersampling at $R{=}10,20,30,40,50$.

Physics-driven \textbf{unrolled networks} that alternate a learned regularizer with k-space data consistency dominate 
accelerated MR reconstruction~\cite{aggarwal2019modl,hammernik2018varnet,sriram2020e2evarnet}, but all require a 
fully-sampled reference for every training example, an acquisition that acceleration exists precisely to avoid. 
Self-supervised methods such as SSDU~\cite{yaman2020ssdu} lift that requirement by splitting the acquired k-space
into an input partition and a held-out consistency target; joint supervised/self-supervised schemes combine the 
two~\cite{yiasemis2025jssl}, keeping supervised accuracy where references exist while still learning from a target dataset 
that has none. Recent CMRxRecon papers pursue the same multi-center generalization goal architecturally, on magnitude-based 
data~\cite{anvari2025genrecmr,tavakoli2025vtransformer}; we target the phase regime instead.

In this work, we propose \textbf{FlowVN-JSSL}, which incorporates a
\textbf{\emph{joint} supervised and self-supervised regime} that supervises velocity phase \emph{directly}. We also design
\textbf{acquisition-robust techniques} including learned coil-map refinement,
coil/noise/phase-ramp augmentation with a \textbf{per-stage acceleration-conditioning
weight} spanning $R{=}10$--$50$ under bounded momentum. Against a training-matched FlowVN
baseline~\cite{vishnevskiy2020deepvn} it improves both magnitude and velocity fidelity at every acceleration, and the
same frozen model transfers to unseen sites and organs without adaptation. Code is
available at \url{https://github.com/cc0231/flowvn_jssl}.

\section{Methods}

\subsection{Problem formulation}
\label{sec:setup}
Per subject, 4D~flow acquires a time-resolved multi-coil k-space volume $k \in
\mathbb{C}^{V\times C\times T\times D\times H\times W}$ with per-coil sensitivity
maps $c$ and a binary sampling mask. The $V=4$ \emph{velocity encodings} are one
flow-compensated reference and three directional encodings; $C$ is the coil count,
$T$ the cardiac frames, and $D\times H\times W$ the volume, in which
$(H,W)=(k_y,k_z)$ is the undersampled phase-encoded plane and $D$ the fully-sampled
readout. The unknown image series $x \in \mathbb{C}^{V\times T\times D\times H\times
W}$ relates to the measurements through the multi-coil forward operator
\begin{equation}
\A(x) = M \odot \mathcal{F}(c \odot x),
\label{eq:fwd}
\end{equation}
with $\mathcal{F}$ the frame-wise spatial DFT along $(k_y,k_z)$ and the mask $M$
applied per encoding and frame; coil combination is the adjoint~$\A^{H}$.

The clinically relevant quantity is not image magnitude but the \textbf{velocity},
carried in the phase difference between each directional encoding and the reference,
\begin{equation}
v_j = \angle\!\big(x_j\,\overline{x_0}\big), \qquad j = 1,2,3,
\label{eq:venc}
\end{equation}
proportional to the blood-velocity component along direction~$j$ through the
encoding velocity (\textsc{venc}). Reconstruction is scored on both magnitude
$|x|$ and the relative phase across encodings, a difference \emph{between}
encodings corrupted by any layer that mixes them (Sec.~\ref{sec:backbones}).
Undersampling is 3-D Cartesian k-t~Gaussian, per encoding, at
$R\in\{10,\dots,50\}$.

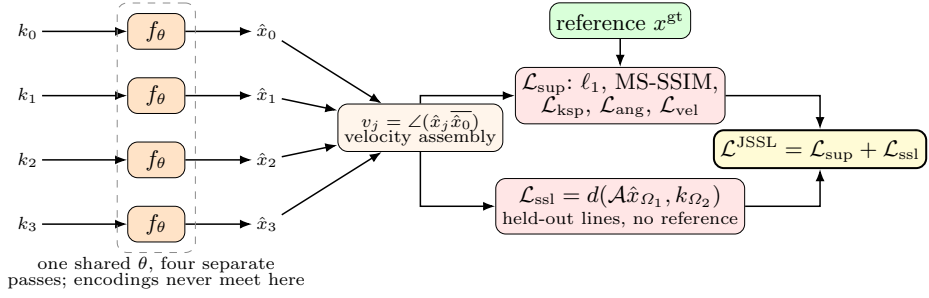
\begin{figure}[t]
\centering
\resizebox{\linewidth}{!}{%
\begin{tikzpicture}[
  font=\footnotesize,
  b/.style   ={draw, rounded corners, inner sep=2.5pt, minimum height=5mm, align=center},
  dat/.style ={b, fill=blue!8},
  gtb/.style ={b, fill=green!15},
  net/.style ={b, fill=orange!22, minimum width=8mm},
  op/.style  ={b, fill=orange!8, font=\scriptsize},
  los/.style ={b, fill=red!10},
  ar/.style  ={-{Latex[length=1.7mm]}, semithick},
  el/.style  ={font=\scriptsize, inner sep=1.5pt},
]
\node[el, anchor=east] (i0) at (0,1.35) {$k_0$};
\node[el, anchor=east] (i1) at (0,0.45) {$k_1$};
\node[el, anchor=east] (i2) at (0,-0.45) {$k_2$};
\node[el, anchor=east] (i3) at (0,-1.35) {$k_3$};
\node[net] (f0) at (1.6,1.35) {$f_\theta$};
\node[net] (f1) at (1.6,0.45) {$f_\theta$};
\node[net] (f2) at (1.6,-0.45) {$f_\theta$};
\node[net] (f3) at (1.6,-1.35) {$f_\theta$};
\node[el] (o0) at (3.15,1.35) {$\hat x_0$};
\node[el] (o1) at (3.15,0.45) {$\hat x_1$};
\node[el] (o2) at (3.15,-0.45) {$\hat x_2$};
\node[el] (o3) at (3.15,-1.35) {$\hat x_3$};
\foreach \i in {0,1,2,3}{\draw[ar] (i\i) -- (f\i); \draw[ar] (f\i) -- (o\i);}
\node[draw=black!45, dashed, rounded corners, fit=(f0)(f3), inner sep=4pt,
      label={[el, align=center]below:one shared $\theta$, four separate\\[-2pt]passes; encodings never meet here}] (SH) {};

\node[op] (vel) at (5.3,0) {$v_j=\angle(\hat x_j\overline{\hat x_0})$\\[-2pt]velocity assembly};
\foreach \i in {0,1,2,3}{\draw[ar] (o\i) -- (vel);}

\node[gtb] (gt)   at (8.1,1.5)  {reference $\xgt$};
\node[los] (lsup) at (8.1,0.45) {$\Lsup$: $\ell_1$, MS-SSIM,\\[-2pt]$\Lksp$, $\Lang$, $\Lvel$};
\node[los] (lssl) at (8.1,-1.1) {$\Lssl=d(\A\hat x_{\Omega_1},k_{\Omega_2})$\\[-2pt]\scriptsize held-out lines, no reference};
\draw[ar] (vel.north) |- (lsup.west);
\draw[ar] (gt)  -- (lsup);
\draw[ar] (vel.south) |- (lssl.west);
\node[b, fill=yellow!20, thick] (tot) at (10.9,-0.3) {$\Ljssl=\Lsup+\Lssl$};
\draw[ar] (lsup) -| (tot.north);
\draw[ar] (lssl) -| (tot.south);
\end{tikzpicture}}
\caption{Method overview. Each velocity encoding $k_j$ passes through the
\emph{same} network $f_\theta$ with $8$ unrolled stages in its own forward.}
\label{fig:method}
\end{figure}

\subsection{Network architecture}
\label{sec:backbones}
Our backbone (Fig.~\ref{fig:method}) is a physics-driven unrolled variational network that alternates a learned 
image-domain regularizer with a data-consistency (DC) update against acquired k-space, over $S=8$ fixed stages 
indexed by~$s$. It processes each velocity encoding independently (\texttt{features\_in}\,$=1$): the four encodings never meet
in a learned layer, so relative velocity phase is fixed by the per-encoding DC
against measured k-space, never altered by cross-encoding mixing. Writing
$\xhat^{(s)}$ for the image estimate at stage~$s$,
\begin{align}
g^{(s)} &= \lambda^{(s)}_{\mathcal{R}}(R)\;\mathcal{R}_{\theta_s}\!\big(\xhat^{(s)}\big)
  \;+\;
  \lambda^{(s)}_{\mathcal{D}}(R)\;
    \A^{H}\!\big(\varphi_{\theta_s}\!\big(\A\xhat^{(s)} - k\big)\big),
\nonumber\\
m^{(s)} &= g^{(s)} + \beta\,m^{(s-1)},\qquad m^{(-1)} = 0,\qquad \beta = \sigma(\alpha),
\nonumber\\
\xhat^{(s+1)} &= \xhat^{(s)} - m^{(s)} .
\end{align}
Here $\A$ is the forward operator~\eqref{eq:fwd}, so the residual $\A\xhat^{(s)}-k$
is supported on acquired samples; $\mathcal{R}_{\theta_s}$ is a sum of orthogonal
3-D convolution branches (spatial and spatio-temporal); and $\varphi_{\theta_s}$ is
a learned pointwise activation on the k-space residual, making the DC update a learned reweighting rather than a plain $\ell_2$ step. 

The recursion carries heavy-ball momentum with a sigmoid-bound coefficient
$\beta=\sigma(\alpha)\in(0,1)$; without the bound the cascade diverged in training. The
stage weights $\lambda^{(s)}_{\mathcal{R}}(R),
\lambda^{(s)}_{\mathcal{D}}(R)$ are learned piecewise-linear functions of~$R$.

\paragraph{Learned coil-map refinement.} A lightweight complex CNN refines the supplied
maps $c$ per coil (coil-count agnostic, renormalized to unit RSS), applied once before
the cascade so every stage's DC uses the corrected maps. Its output layer is
zero-initialized, so training starts from the unrefined maps and departs only as the data
warrant, absorbing site- and array-dependent error rather than committing to a fixed
ESPIRiT~\cite{uecker2014espirit} estimate.

\subsection{Training objective and evaluation}
\label{sec:jssl}
Joint supervised and self-supervised training~\cite{yiasemis2025jssl} trains one
network from data with and without references simultaneously, lowering estimator
variance relative to self-supervision alone. It optimizes the unweighted sum
$\Ljssl = \Lsup + \Lssl$, each term computed over a random sample from an
independent, mutually exclusive dataset.

\paragraph{Evaluation metrics.}\phantomsection
\label{sec:eval}
Magnitude accuracy is scored by normalized root mean squared error (nRMSE) and
multi-scale structural similarity (MS-SSIM); velocity accuracy by
\begin{equation}
\mathrm{RelErr} = \frac{\big\lVert\,|v^{gt}|-|\hat v|\,\big\rVert_2}{\big\lVert\,|v^{gt}|\,\big\rVert_2}
\qquad
\mathrm{AngErr} =  \arccos\!\left(\frac{\langle \hat v, v^{gt}\rangle}{\lVert\hat v\rVert\,\lVert v^{gt}\rVert}\right)
  \cdot\frac{180}{\pi},
\end{equation}
where $v=(v_1,v_2,v_3)$ is the velocity vector.

All metrics are computed inside the segmentation mask $\Omega\equiv\Omega_{\mathrm{seg}}$
and averaged over encodings and frames, both volumes first scaled by
$\max_{\Omega}|\xgt|$.

\paragraph{Supervised stream.}\phantomsection\label{sec:gt} The reconstruction is
penalized inside $\Omega_{\mathrm{seg}}$ by
\begin{equation}
\Lsup = \lVert \xhat - \xgt \rVert_1
      + \lambda_{\mathrm{ssim}}\,(1-\mathrm{MS\text{-}SSIM})
      + \lambda_{\mathrm{ksp}}\,\Lksp
      + \lambda_{\mathrm{ang}}\,\Lang
      + \lambda_{\mathrm{vel}}\,\Lvel .
\end{equation}

The direction term
$\Lang = 1 - \langle v^{\mathrm{pred}}, v^{\mathrm{gt}}\rangle \mathbin{/}
(\lVert v^{\mathrm{pred}}\rVert\,\lVert v^{\mathrm{gt}}\rVert)$ penalizes the angle
between predicted and GT velocity vectors $v=(v_1,v_2,v_3)$~\eqref{eq:venc}; the
vector term $\Lvel$ penalizes the wrapped per-component error
$\angle\big(e^{\mathrm{i}v^{\mathrm{pred}}_j}\,\overline{e^{\mathrm{i}v^{\mathrm{gt}}_j}}\big)$,
correct across the $\pm\pi$ boundary and reducing RelErr and AngErr together rather
than direction alone. Both terms are weighted by ground-truth velocity
magnitude so noise-dominated low-velocity voxels do not contaminate the gradient.

Both k-space terms use one normalized, reference-free discrepancy between a prediction and measured lines,
\begin{equation}
d(a,b)=\tfrac{1}{2}\,\frac{\lVert a-b\rVert_2}{\lVert b\rVert_2}
      +\tfrac{1}{2}\,\frac{\lVert a-b\rVert_1}{\lVert b\rVert_1},
\label{eq:disc}
\end{equation}
scale-invariant across the multi-vendor cohort. The supervised stream applies it to
the sample's own acquired lines, $\Lksp=d(\A\xhat,k)$, at no reference cost.

\paragraph{Self-supervised stream.} Following SSDU~\cite{yaman2020ssdu}, the acquired
mask is split into an input partition $\Omega_1$ and a disjoint loss partition
$\Omega_2$ ($\rho=0.2$ of acquired points); the network reconstructs from $\Omega_1$
and consistency on the held-out lines gives
$\Lssl=d(\A\xhat_{\Omega_1},k_{\Omega_2})$, requiring no reference. Since $\Omega_2$
shrinks with $R$ we report metrics per~$R$; being per-encoding, the stream draws one
random encoding per step. The reference-free samples are training
subjects whose references are withheld from the model throughout training.

\paragraph{Acquisition augmentation and normalization.}\phantomsection
\label{sec:robust}
Three random augmentations are applied to the \emph{supervised} stream: k-space
\textbf{noise} scaled to the sampled-k-space RMS, multiplicative
\textbf{coil-map jitter} (clean-map target, teaching tolerance to coil-map error),
and a linear readout \textbf{phase ramp} (target inherits it, keeping the task
well-posed). Each needs a fully-sampled source or true-map target the self-supervised
stream lacks. Separately, inputs are scaled by a k-space RMS with the top $0.5\%$ of
magnitudes winsorized, stabilizing scale against vendor-specific high-magnitude
outlier samples; this normalization applies to both streams.

\section{Experiments and Results}
\paragraph{Data and splits.} The reconstruction task provides an aortic cohort of 138
subjects from 6 center/scanner domains (1.5 and 3\,T, multiple vendors). The
\textbf{self-supervised stream} takes all 6 subjects of the Center011 cohort. The remaining
$119$ subjects form the \textbf{supervised stream}, and $13$ subjects are \textbf{held out}
for local evaluation, with the three groups disjoint at the subject level. Subjects acquired
on the under-represented Philips systems (Center010 and Center012) were sampled twice per
epoch to counter the predominance of GE data in the cohort. See supplementary file for
detailed subject information.

\paragraph{Implementation.} The network was implemented in Python 3.10.9 with
PyTorch 2.3.1 and PyTorch Lightning 2.2.0, and trained on a single NVIDIA RTX\,6000 Pro
GPU. The learning rate started at $5\times10^{-4}$ and was reduced by a factor of $0.7$
after two epochs without improvement; the batch size was 1, gradients were clipped at
norm 1.0, and the random seed was fixed at 42. A loss-spike guard with a threshold of 8
reverted non-finite or extreme-loss batches to the last clean snapshot. All configurations
were trained for a fixed budget of 16 epochs and evaluated at the final epoch, so no
checkpoint was selected using held-out data. The loss weights were chosen empirically as
$\lambda_{\mathrm{ssim}}=0.16$, $\lambda_{\mathrm{ksp}}=0.3$ and
$\lambda_{\mathrm{ang}}=\lambda_{\mathrm{vel}}=1.0$, with $\lambda_{\mathrm{ang}}$
ramped linearly to its full value over the first four epochs.

\paragraph{Network.} The network used 8 unrolled stages and $320{,}761$ trainable
parameters. Each stage's regularizer used 3-D convolutions with kernel size 7, one input
feature per encoding and 24 output features, and the learned k-space activation was
parameterized by 71 weights.

\paragraph{Augmentation.} Augmentation strengths were $0.1$ for coil-map jitter, $0.04$
for additive noise relative to the sampled k-space RMS, and $1.0$ for the readout phase
ramp; all three were applied to the supervised stream only. Inputs were scaled by a
k-space RMS with the top $0.5\%$ of magnitudes winsorized, a normalization applied to
both streams.

\paragraph{Baselines.} The FlowVN baseline~\cite{vishnevskiy2020deepvn} was trained on
the same split and hyperparameters as the proposed model. As a classical comparison we
used compressed-sensing, locally low-rank (CS-LLR)
reconstruction~\cite{lustig2007sparsemri,zhang2015llr}, solved by proximal gradient
descent for 70 iterations ($\lambda_{\mathrm{LLR}}{=}0.25$, $\alpha{=}0.3$,
$\tau{=}0.97$) at every~$R$ with the total-variation term disabled.

\subsection{Quantitative and qualitative results}
\begin{figure}[tbp]
\centering
\includegraphics[width=1.0\linewidth]{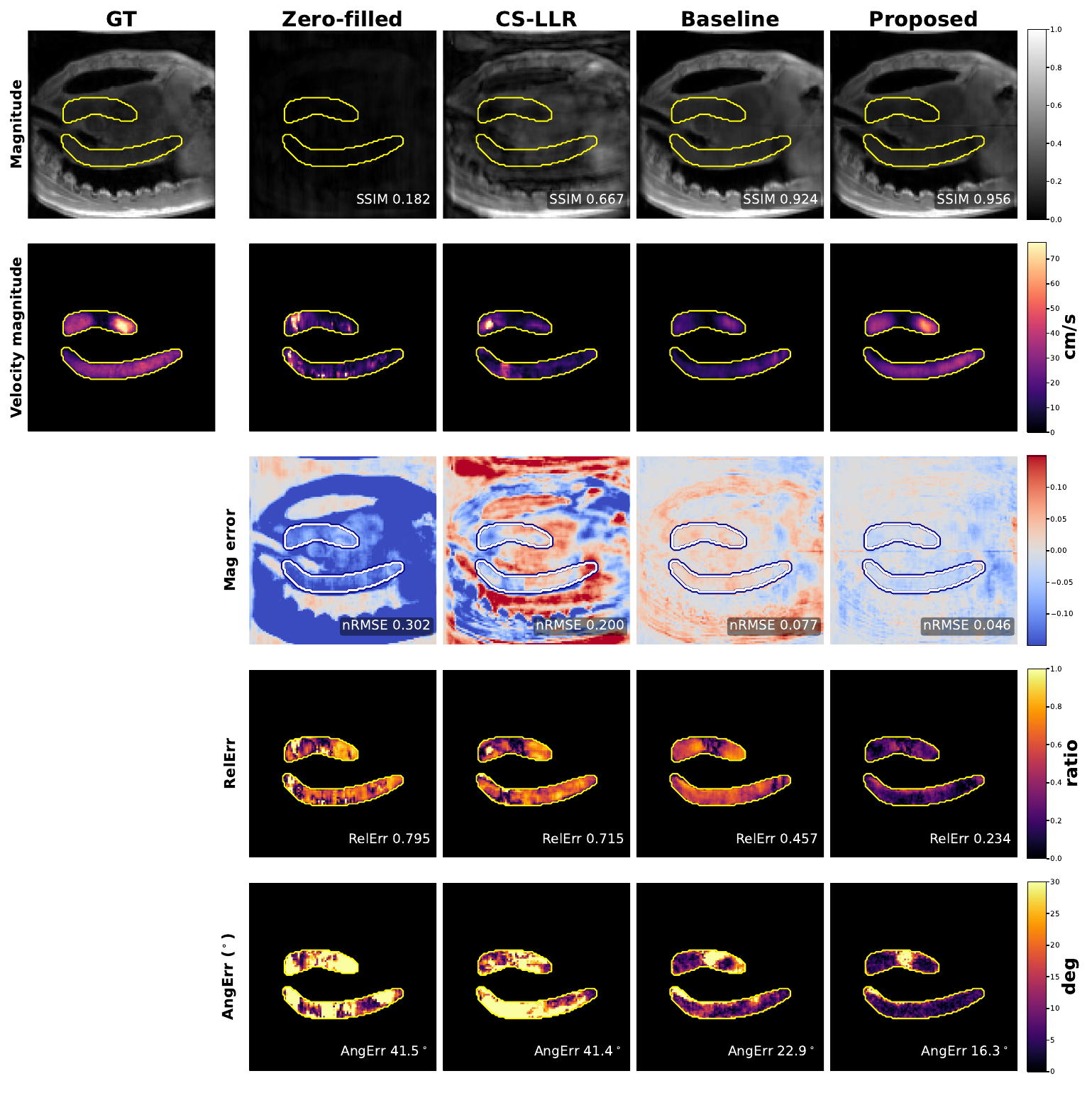}
\caption{Qualitative reconstruction comparisons of four approaches at $R{=}30$ for a held-out subject of a peak-systole frame. ROI of mask was labeled on each panel.}
\label{fig:recon}
\end{figure}

\begin{figure}[tbp]
\centering
\includegraphics[width=0.8\linewidth]{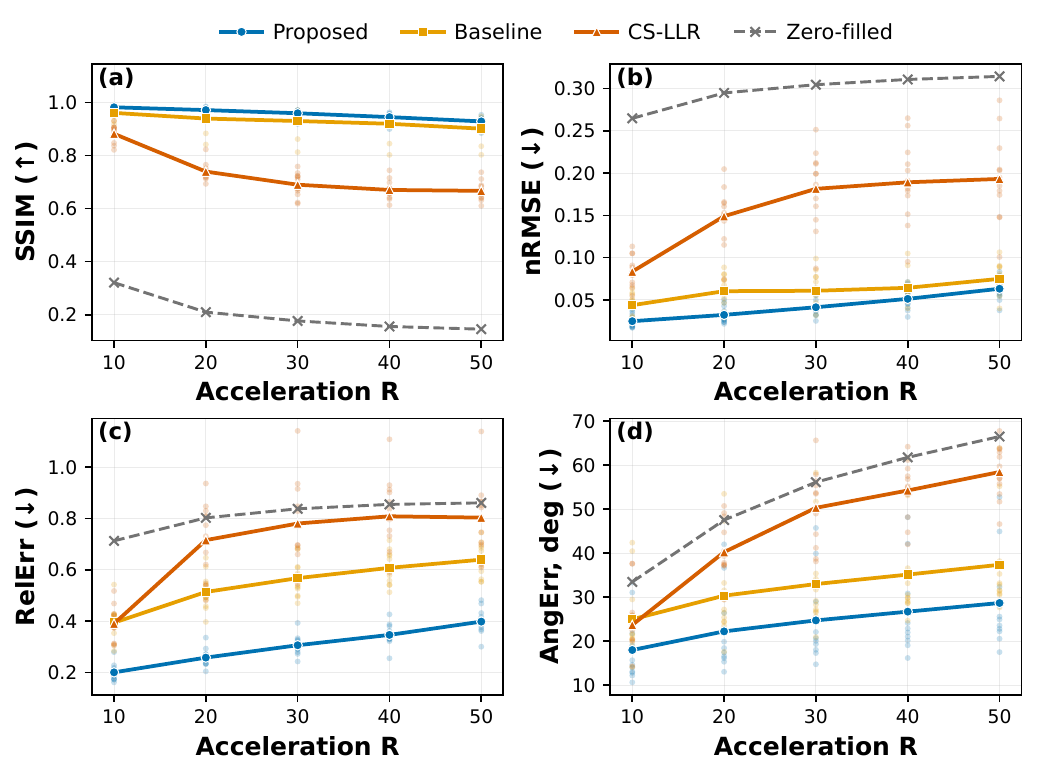}
\caption{Local held-out accuracy vs.\ acceleration $R$ of different approaches. Faint points are values on individual
held-out subjects; solid lines their per-$R$ means.}
\label{fig:rsweep}
\end{figure}

Fig.~\ref{fig:recon} shows one held-out case at $R{=}30$ reconstructed with four approaches (zero-filled, CS-LLR, baseline, proposed method): the baseline and proposed networks are visually comparable in magnitude, with the proposed model showing slightly lower magnitude error, while it is visibly better on the velocity-based metrics in the AngErr and RelErr maps. Fig.~\ref{fig:rsweep} reports magnitude and velocity metrics across acceleration factors for the compared reconstruction approaches. CS-LLR degrades sharply once $R$ exceeds~$10$, while both networks worsen gradually.
Between the two learned methods, SSIM is $0.021$--$0.032$ higher at every~$R$,
whereas RelErr is $38$--$50\%$ lower for the proposed model, and the angular error margin widens monotonically with 
acceleration from $7.0^\circ$ at $R{=}10$ to $8.7^\circ$ at $R{=}50$. The proposed model is also more accurate on
\emph{every} held-out subject at every~$R$ on both velocity metrics.

\textbf{Generalization to unseen sites and organs.}\phantomsection\label{sec:gen}
Table~\ref{tab:main} evaluates the \emph{same} frozen model on the challenge's
cross-site and cross-organ sets with no retraining or per-domain fine-tuning. We compare our proposed model with a supervised twin, trained without the joint objective, and the FlowVN baseline. Against the baseline the proposed model gains $0.063$ SSIM, lowers RelErr by $31\%$ and lowers AngErr by $17.7^\circ$, so the local margin of $7.0$--$8.7^\circ$ roughly
doubles once the site and scanner are unseen. Against the supervised twin the margin is much smaller, though the proposed model still leads on every metric. Cross-organ, the proposed model leads on every metric and the angular margin of $7.16^\circ$ is close to the local held-out range, so the advantage carries to unseen anatomy at roughly its in-domain scale.

\begin{table}[!htbp]
\centering
\caption{Zero-shot generalization of different approaches on two out-of-distribution datasets: cross-site (an unseen site and scanner model) and cross-organ (non-aortic, categorized by organ).}
\label{tab:main}
\footnotesize\setlength{\tabcolsep}{3.4pt}\renewcommand{\arraystretch}{0.95}%
\begin{tabular}{lcccc}
\toprule
\emph{Generalization (frozen model, zero-shot)}
& SSIM $\uparrow$ & nRMSE $\downarrow$ & RelErr $\downarrow$ & AngErr ($^\circ$) $\downarrow$ \\
\midrule
Cross-site, \textbf{proposed} & \textbf{0.942} & \textbf{0.084} & \textbf{0.467} & \textbf{28.98} \\
Supervised twin (no joint objective) & {0.938} & {0.086} & {0.488} & {29.39} \\
FlowVN baseline & {0.879} & {0.123} & {0.677} & {46.67} \\
\midrule
Cross-organ \emph{mean}, \textbf{proposed}  & \textbf{0.966} & \textbf{0.052} & \textbf{0.398} & \textbf{29.51} \\
\quad \emph{Carotid} & {0.961} & {0.055} & {0.247} & {18.83} \\
\quad \emph{Cerebrovascular} & {0.950} & {0.046} & {0.470} & {32.81} \\
\quad \emph{Portal Vein} & {0.975} & {0.058} & {0.402} & {32.71} \\
\quad \emph{Renal Artery} & {0.976} & {0.050} & {0.472} & {33.71} \\
Cross-organ \emph{mean}, FlowVN baseline & {0.957} & {0.065} & {0.473} & {36.67} \\
\bottomrule
\end{tabular}
\end{table}

\subsection{Ablation studies}
Table~\ref{tab:abl} reports an incremental ablation, adding one mechanism at a time. Acceleration conditioning
and the dual-domain $\Lksp$ term are the components that reduce the velocity metrics on
their own, together giving around a $10\%$ reduction in RelErr and $3^\circ$ of AngErr against the
baseline, the latter trading magnitude for velocity accuracy.

Therefore, JSSL is the largest single increment on every velocity metric and the only one
that changes the hardest regime. It reduces RelErr from $0.501$ to $0.376$ and AngErr from
$33.04^\circ$ to $28.18^\circ$ while also raising SSIM by $0.013$, so the velocity gain is
not bought with magnitude fidelity. The separation is sharpest under the heaviest
undersampling: rows~2--4 leave AngErr$_{R50}$ within $0.5^\circ$ of the baseline, whereas
JSSL drops it to $35.01^\circ$. As the architecture of JSSL is targeted for higher accuracy at cross-domain data without reference and re-training by definition, it is impossible to isolate each stream and strictly exploit the architectural advantage during ablation experiments.

\begin{table}[!htbp]
\centering
\caption{ Ablation studies. Step sizes, optimizer and scheduler are fixed across configurations. Runs are single-seed (seed 42).}
\label{tab:abl}
\footnotesize\setlength{\tabcolsep}{3.5pt}\renewcommand{\arraystretch}{0.95}%
\begin{tabular}{lcccc}
\toprule
Configuration & SSIM $\uparrow$ & RelErr $\downarrow$ & AngErr$_{\mathrm{mean}}$ $\downarrow$ & AngErr$_{R50}$ $\downarrow$\\
\midrule
Baseline + bounded momentum & {0.951} & {0.552} & {36.25$^\circ$} & {43.87$^\circ$} \\
\quad + acceleration conditioning & {0.955} & {0.509} & {34.00$^\circ$} & {43.96$^\circ$} \\
\quad + acquisition-robust techniques & {0.956} & {0.507} & {34.18$^\circ$} & {43.40$^\circ$} \\
\quad + dual-domain $\Lksp$~\eqref{eq:disc} & {0.946} & {0.501} & {33.04$^\circ$} & {43.50$^\circ$} \\
\quad + JSSL and velocity losses & \textbf{0.959} & \textbf{0.376} & \textbf{28.18$^\circ$} & \textbf{35.01$^\circ$} \\
\bottomrule
\end{tabular}
\end{table}

\FloatBarrier

\section{Discussion and Conclusion}
\paragraph{Where the gains fall.} Results that the two learned methods differ almost
entirely on velocity (Figs.~\ref{fig:recon},~\ref{fig:rsweep}) are by construction,
not incidental: the diagnostic signal of 4D~flow lives in the \emph{relative phase}
across the four encodings~\eqref{eq:venc}, not in image magnitude, which saturates
early for both models and leaves little SSIM headroom. The
better-supported mechanism is that $\Lang$ and the wrapped, $\pm\pi$-correct $\Lvel$
supervise relative phase \emph{directly}, weighted by ground-truth velocity
magnitude so noise-dominated voxels do not contaminate the gradient, since per-encoding
processing alone shows no velocity gain.

\paragraph{Sites vs.\ organs.} The frozen model transfers better across anatomy than
across acquisition systems (Table~\ref{tab:main}), and the two shifts are not alike. A new
anatomical region changes the \emph{content} the network sees but not the measurement
model, so data consistency stays valid and the learned regularizer still applies. A new
site changes the acquisition itself: under-represented vendors perturb the sampling and
intensity statistics the physics operators rely on, so the cross-site loss concentrates in
magnitude and RelErr, while angular error is essentially unchanged.

\paragraph{Limitations.} Splitting an already-undersampled acquisition leaves the
consistency loss only $\approx\!0.4\%$ of k-space at $R{=}50$, bounding reference-free
supervision where accuracy is hardest. The final ablation increment does not fully isolate the joint objective: it bundles JSSL with
the velocity-specific losses and simultaneously adds the reference-free
stream, so some variables are not strictly controlled. A controlled attribution is left to future work for better quantitative evaluations. Supervised
inputs are \emph{retrospectively} undersampled. RelErr and AngErr are voxel-wise
proxies for the derived clinical quantities, and background phase is left to standard
downstream correction.

\paragraph{Conclusion.} We incorporated a joint training regime into a 4D flow reconstruction network with direct velocity-phase supervision. It shows outstanding capability in reconstructing 4D flow images across various acceleration factors, and the same frozen model transfers to unseen sites and organs without adaptation.

\begin{credits}
\subsubsection{\discintname}
The authors have no competing interests to declare.
\end{credits}

\FloatBarrier
\bibliographystyle{splncs04}
\bibliography{references}

\end{document}